\documentclass[showpacs,prb,twocolumn,floats,superscriptaddress]{revtex4}

\usepackage{graphicx}
\usepackage{amsmath}
\usepackage{amssymb}
\usepackage[colorlinks=true,citecolor=blue,linkcolor=blue]{hyperref}
\usepackage{amsfonts}
\usepackage[latin9]{inputenc}
\usepackage{color}

\DeclareMathOperator{\arsinh}{arsinh}    
\DeclareMathOperator{\arcosh}{arcosh}

\DeclareMathOperator{\Tr}{Tr}

\begin{document}

\title{Spin and valley  polarization  of plasmons in silicene due to external fields}
\date{\today }
\author{B. Van Duppen}
\email{ben.vanduppen@uantwerpen.be}
\affiliation{Department of Physics, University of Antwerp, Groenenborgerlaan 171, B-2020 Antwerp, Belgium}
\author{P. Vasilopoulos}
\affiliation{Concordia University, Department of Physics, 7141 Sherbrooke Ouest Montr\'{e}al, Qu\'{e}bec H4B 1R6, Canada}
\author{F. M. Peeters}
\affiliation{Department of Physics, University of Antwerp, Groenenborgerlaan 171, B-2020 Antwerp, Belgium}

\begin{abstract}
The electronic properties of the novel two dimensional (2D) material silicene are strongly influenced by the application of a perpendicular electric field  $E_z$ and of an exchange field $M$ due to adatoms positioned on the surface  or a ferromagnetic substrate. Within the random phase approximation, we investigate how electron-electron interactions are affected by these fields and present analytical and numerical results for the dispersion of plasmons, their lifetime, and their oscillator strength.  We find that the combination of the fields  $E_z$ and $M$ brings a spin and valley texture to the particle-hole excitation spectrum and  allows the formation of spin- and valley-polarized plasmons. 
When the Fermi level lies in the gap of one spin in one valley, the intraband region of the 
corresponding spectrum disappears. For zero $E_z$ and  finite $M$ the spin symmetry is broken  and spin polarization is possible.
The lifetime and oscillator strength of the plasmons are shown to depend strongly on the number of spin and valley type electrons that  form the  electron-hole pairs.
\end{abstract}

\maketitle

\section{Introduction}

Since its realization as a truly two-dimensional (2D) material,
graphene has attracted much interest, both due to fundamental science and
technological importance in various fields \cite{1}. However, the realization of a
tunable band gap, suitable for device fabrications, is still challenging and spin-orbit-coupling (SOC) is very
weak in graphene. To overcome these limitations researchers have been increasingly studying
similar materials. One such material, called silicene,  is a monolayer honeycomb structure of silicon and has been predicted to be stable \cite{xx}.
Already several attempts have been made to synthesize it \cite{Feng2012, Vogt2012, Fleurence2012,YanVoon2014} and its properties are reviewed in Ref. \onlinecite{Kara2012}.

Despite controversy over whether silicene has been created or not  \cite{zz}, it is expected to be  
an excellent candidate material because it has a strong SOC and an electrically tunable
band gap \cite{4,5,9}. It's  a single layer of silicon atoms with
a honeycomb lattice structure and compatible with silicon-based  electronics that dominates the
semiconductor industry. Silicene has Dirac cones similar to those of graphene and density functional
calculations showed that the SOC induced  gap in it is about 1.55 meV 
\cite{4,5}.  Moreover,  very recent theoretical studies predict  the  stability of  silicene 
on non metallic surfaces such as graphene  \cite{yon}, boron nitride or SiC \cite{liu1}, and in graphene-silicene-graphene structures \cite{meh}. 
Besides the strong SOC, another salient feature of silicene is its buckled lattice structure with the A and B sublattice planes separated by a vertical distance $2\ell$ so that inversion symmetry can be broken by an external perpendicular electric field resulting in a staggered potential \cite{9}. Accordingly, the energy gap  in it can be controlled electrically. Due to this unusual band structure, silicene is expected to show exotic properties such as quantum spin/valley and anomalous Hall effects \cite{9,10, Tabert2013}, magneto-optical and electrical transport \cite{13}, etc.. 

One  interesting property of 2D materials is their use in developing fast plasmonic devices \cite{Grigorenko2012}. Plasmons are quantized collective oscillations of the electron liquid. In graphene they have been studied extensively both theoretically \cite{theor} and experimentally \cite{exp}. So far though in silicene the relevant studies are limited \cite{Tabert2014a, Chang2014a} and do not take into account the effect of an exchange field $M$  which can be induced by ferromagnetic adatoms\cite{Qiao2010} or a ferromagentic substrate\cite{Haugen2008, Yokoyama2014}. This is important as this field  leads to spin- and valley-polarized currents\cite{Yokoyama2013} and,  as will be shown, brings a spin and valley texture to the particle-hole excitation spectrum (PHES).  

The application of a perpendicular electric field $E_z$ enhances the SOC gap for one spin  and valley type, while it shrinks it for the other \cite{9, Drummond2012}. Together with the influence of the field M, which breaks the  spin degeneracy, this leads to valley polarization and the occurrence of the anomalous Hall effect \cite{9}. In this work we combine these two peculiar features and calculate the dynamical polarization function  within the random phase approximation (RPA) and silicene's plasmonic response to optical excitations. We investigate how the  fields  $E_z$ and $M$ influence such a response and calculate the decay rate and oscillator strength (not  evaluated in Ref.  \onlinecite{Tabert2014a}) of the plasmons.

In Sec. \ref{Sec:Electrons} we discuss the one-electron Hamiltonian, the energy spectrum, and the density of states as well as their dependence on the SOC. Then we present analytical results for the polarization in Sec. \ref{Sec:Polarization} and use them to calculate and discuss the plasmon dispersion and corresponding decay rate and oscillator strength in Sec. \ref{Sec:NumericalRes}. In Sec. \ref{Sec:Concl} we make concluding remarks.

\section{Basic formalism}\label{Sec:Electrons}

Silicene consists of a hexagonal lattice of silicon atoms.\cite{Kara2012, 4, 5} Similar to graphene, the  silicon atoms make up two trigonal sublattices which we call $A$ and $B$ sublattices. These sublattices are vertically displaced by\cite{9} $2\ell = 0.46$\AA,  and form  the buckled structure of silicene. Due to the large ionic radius, the interatomic distance of silicene is also larger than that of graphene, measuring $a=3.89$\AA, whereas for graphene this is $a=1.42$\AA. Because of the buckling, the conduction electrons move in a hybridisation\cite{xx} of the $p_{z}$ orbitals with the $\sigma$ orbitals and therefore the SOC is strong and cannot be neglected.\cite{Tahir2013}

The behaviour of the electrons in silicene can be described using a four-band next-nearest-neighbour tight- binding model. Near the $K_{\eta }$ point, the corresponding Hamiltonian is given by \cite{4, 9}

\begin{eqnarray}
H_{\eta } &=&\hbar v_{F}\left( \eta k_{x}\tau _{x}+k_{y}\tau _{y}\right)
+\eta \tau _{z}h_{11}-\ell E_{z}\tau _{z}  \notag \\
&&+M\sigma _{z}+\lambda _{R1}\left( E_{z}\right) \left( \eta \tau _{x}\sigma
_{y}-\tau _{y}\sigma _{x}\right) /2,
\label{HamStart}
\end{eqnarray}
where $h_{11}=\lambda _{SO}\sigma _{z}+a\lambda _{R2}\left( k_{y}\sigma_{x}-k_{x}\sigma _{y}\right) $. Here  $v_{F}\approx 0.5 \times 10^{6}m/s$ as the Fermi velocity of the electrons \cite{4}, $\boldsymbol{k} =\left( k_{x},k_{y}\right) $  the 2D wave vector, $\eta $ distinguishes between the two valleys, and $\lambda_{SO}\approx 3.9$ meV is the intrinsic SOC strength. Further, $\lambda_{R1}( E_{z}) $ and $ \lambda _{R2}\,$  represent the Rashba SOC due to the external electric field $E_{z}$ and  the intrinsic Rashba SOC which is present due to the buckling of silicene\cite{5, Min2006}, respectively.

The exchange field  $M$ can be induced by ferromagnetic adatoms or a ferromagnetic substrate.  It's value is predicted to be $M\approx 3$ meV for graphene deposited on a EuO substrate\cite{Haugen2008}. The exchange effect is due to the proximity of the $Eu^{2+}$ moments and is therefore tunable by varying the distance between the substrate and the silicene layer. In this paper we will use a slightly bigger value of $M$ to make its effect more clear. The Pauli matrices $\sigma_{i}$ and $\tau _{i}$ correspond to the physical spin and the sublattice pseudospin, respectively.

One can write this Hamiltonian in the basis of atomic spin-orbital eigenfunctions of the sublattices $A$ and $B$ for both spin components,  $\Psi =\left\{ \psi _{A\uparrow },\psi _{B\uparrow },\psi_{A\downarrow },\psi _{B\downarrow }\right\}^{T} $, near the point $K_{+}$ as ($k_{\pm } =k_{x}\pm ik_{y}$)

\begin{equation}
H_{+}=\left[ 
\begin{array}{cccc}
E\left( 1,1\right) & \hbar v_{F}k_{-} & ia\lambda _{R2}k_{-} & 0 \\ 
\hbar v_{F}k_{+} & E\left( 1,-1\right) & i\lambda _{R1} & -ia\lambda
_{R2}k_{-} \\ 
-ia\lambda _{R2}k_{+} & -i\lambda _{R1} & E\left( -1,1\right) & \hbar
v_{F}k_{-} \\ 
0 & ia\lambda _{R2}k_{+} & \hbar v_{F}k_{+} & E\left( -1,-1\right)
\end{array}
\right],  \label{HamTotal}
\end{equation}
where $E\left( s_{z},t_{z}\right) =\Delta_{s_z}t_{z}+Ms_{z}$ and $\Delta_{s_{z}}=s_{z}\lambda _{SO}-\ell E_{z}$.

\subsection{Spin-orbit interaction}
\begin{figure}[tb]
\centering
\vspace{0.4cm}
\includegraphics[width= 8.5cm]{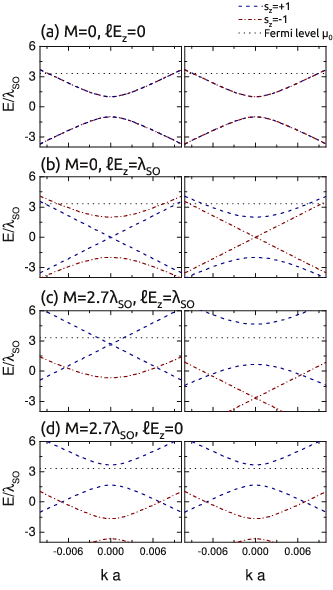}
\vspace{-0.3cm}
\caption{(Colour online)  Energy spectrum of silicene  
versus wave vector $k$ in the $K_{+}$  (left column) and the  $K_{-}$ valleys (right column) for different values of the  electric $E_{z}$ and exchange  $M$ fields, as specified, and $\lambda_{SO}=3.9$ meV.  The blue dashed curves pertain to $s_{z}=+1$ electrons and the red dash-dotted curve 
to $s_{z}=-1$ electrons when the terms $
\lambda_{R2}$ are neglected.  The   dotted lines show the Fermi level  $ \mu_{0}=13$ meV. This  $ \mu_{0}$ value will be used in the entire paper.}
\label{BandPlot}
\end{figure}

In Fig. \ref{BandPlot} we show the energy spectrum for $\lambda_{SO}=3.9$ meV and several combinations of   $E_z$ and $M$ values  near both valleys. In the Hamiltonian $\left( \ref{HamStart}\right) $ three types of SOC are included. One is  represented by the parameter $\lambda_{SO}$ and introduces a gap in the spectrum as shown in Fig. \ref{BandPlot}(a). It is diagonal in the spin basis, so both spin components are treated equally. The   terms $\lambda_{R1}$ and $\lambda_{R2}$ do mix up the spin states. However, as  shown below, the effect of these terms is very small and  the spin remains an approximate good quantum number.

The  term $\lambda _{R1}\left(E_{z}\right)$ describes the Rashba SOC due to the field $E_{z}$. Note, however, that for a wide range of $E_{z}$ values this term can be safely neglected since at fields of the order of the critical field $E_{c}=\lambda_{SO}/\ell$, it's value is approximately $\lambda _{R1} \approx 10^{-3}\lambda_{SO}$.\cite{9, 4, Min2006}

A unitary transformation of Eq. $\left( \ref{HamTotal}\right) $ allows more insight into the importance of $\lambda _{R2}$. With the combinations 
\begin{eqnarray}
\psi _{1} &=&[\hbar v_{F}\psi _{B\uparrow }+ia\lambda _{R_{2}}\psi
_{A\downarrow }]/\hbar v_{F}^{\prime }, \\
\psi _{2} &=&[\hbar v_{F}\psi _{A\downarrow }+ia\lambda _{R2}\psi
_{B\uparrow }]/\hbar v_{F}^{\prime },
\end{eqnarray}
where $\hbar v_{F}^{\prime }=\hbar v_{F}(1+\xi^{2})^{1/2}$ and $\xi = a\lambda _{R2}/\hbar v_{F}\approx 0.5\times 10^{-3}$, the new basis $\Psi_{s_{z}}=\left\{ \psi _{A\uparrow },\psi _{1},\psi _{2},\psi_{B\downarrow }\right\}^{T} $ transforms Eq. (\ref{HamTotal}) into the form
\begin{widetext}
\begin{equation}
H_{+}^{\prime }=\left[ 
\begin{array}{cccc}
E\left( 1,1\right) & \hbar v_{F}^{\prime }k_{-} & 0 & 0 \\ 
\hbar v_{F}^{\prime }k_{+} & E\left( 1,-1\right) -2M'\xi^{2} & -2iM' c & 0 \\ 
0 & 2iM' c& E\left(-1,1\right) +2M'\xi^{2} & \hbar v_{F}^{\prime }k_{-}\\ 
0 & 0 & \hbar v_{F}^{\prime }k_{+} & E\left( -1,-1\right)
\end{array}
\right] ,
\end{equation}
\end{widetext}
where $M^{\prime }=\ell E_{z}+M$ and $c=\xi/(1+\xi^{2})^{1/2}$. 

The effect of $\lambda _{R2}$ is thus threefold. To first order in $\xi$ it induces a change in the Fermi velocity, $v_{F}\rightarrow v_{F}^{\prime }$, and it couples the spin components by virtue of a finite electric $E_{z}$ or exchange $M$ field. Additionally, it affects the gap to second order in $\xi$ by a diagonal term that is linear in $E_{z}$ and $M$. However, these effects are very small due to the smallness  of $\xi$. We shall therefore neglect $\lambda _{R2}$.

\subsection{Effective Hamiltonian}

The  approximations referred to above decouple the two spin states.  Because the valleys are independent, we can describe electrons in silicene as particles that have a spin- and valley-dependent gap. In the basis  $\Psi_{\eta s_{z}}=\left\{ \psi _{A,\eta s_{z}},\psi _{B,\eta s_{z}}\right\}^{T} $ the Hamiltonian becomes

\begin{equation}
H_{\eta}^{s_{z}}=\left[ 
\begin{array}{cc}
\Delta _{\eta s_{z}} +s_{z}M & \hbar v_{F}\eta k^{\eta}_{-} \\ 
\hbar v_{F}\eta k^{\eta}_{+} & -\Delta _{\eta s_{z}} +s_{z}M
\end{array}
\right] ,
\label{Eq:Hamiltonian2By2}
\end{equation}
where $s_{z}=\pm 1$ is the spin quantum number, $\eta$ denotes the valley and $k^{\eta}_{\pm}$ equals $k_{\pm}$ for $\eta=+1$ and $k^{*}_{\pm}$ for $\eta=-1$. 

The gap is independent of $M$ and given by 
\begin{equation}
2|\Delta _{\eta s_{z}}|=2|\eta s_{z}\lambda _{SO}-\ell E_{z}|.
\label{Eq:Gap}
\end{equation}
  
Equation (\ref{Eq:Hamiltonian2By2}) corresponds to a 2D Dirac Hamiltonian of particles with mass $\Delta_{\eta s_{z}}$. Note that in graphene because of the low buckling, both the SOC and the effect of an electric field on the gap  are negligible.\cite{Min2006} 
The energy spectrum obtained from Eq. (\ref{Eq:Hamiltonian2By2}) reads 
\begin{equation}
E_{\eta }^{s_{z}}=s_{z}M+\lambda [\hbar^2 v_{F}^2k^{2}+\Delta _{\eta s_{z}}^2]^{1/2},
\end{equation}
where $\lambda=1 (-1)$ denotes the electron (hole) states. We show it for all spin and valley components in Fig. \ref{BandPlot} for different values of $\ell E_z$ and $M$. 
In Fig. \ref{BandPlot}(b) $\ell E_z$ attains its critical value $E_{c}=\lambda _{SO}/\ell$, for which the $s_{z}=+1$ spin component  has a linear massless dispersion, while the $s_{z}=-1$ one has a large  gap  near the $K_{+}$ point. In Fig. \ref{BandPlot}(c) the field $M$ is  finite; it  displaces both spin components in opposite directions  and results in a spectrum that is different for each spin and valley type of electron. In Fig. \ref{BandPlot}(d) only the $M$ field is present and leads to spin polarization but the valleys remain equivalent. 

\begin{figure}[tb]
\centering
\includegraphics[width= 8.5cm]{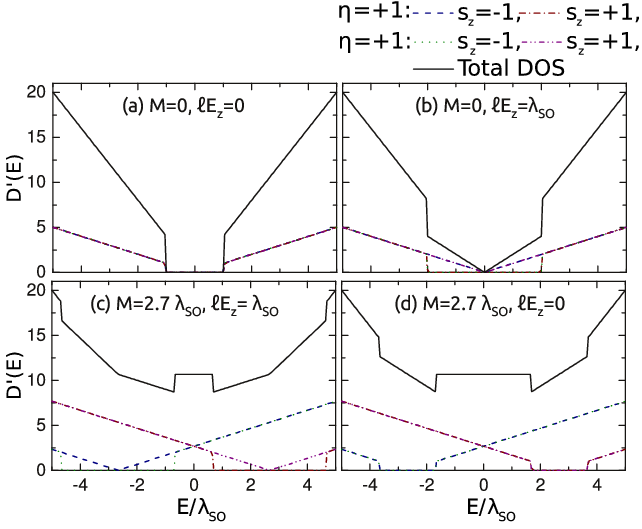}
\caption{(Colour online) DOS in silicene for different values, specified, of the electric $E_z$ and exchange $M$ fields. The solid black curve is the total DOS. The other curves correspond to different valleys and spins as shown at the top of the figure. The DOS is expressed as a dimensionless quantity defined by $D^{\prime}(E)=(2\pi \hbar^{2} v_{F}^{2}/\lambda_{SO})\,D(E)$.}
\label{Fig:DOS}
\end{figure}

This energy spectrum gives rise to a density of states $D(E)$ (DOS) with a structure that depends sensitively on the values of the electric $E_z$ and exchange $M$ fields. With $\bar{E}= E-s_{z}M$ the full expression for $D( E)$ is \cite{Stille2012} 
\begin{equation}
D\left( E\right) =\sum_{\eta =\pm 1}\sum_{s_{z}=\pm 1}\frac{\left \vert \bar{E}\right \vert }{2\pi \hbar^2 v_{F}^{2}}\,\Theta \big
( \left \vert \bar{E}\right \vert -\left \vert \Delta _{\eta s_{z} }\right \vert\big) .
\end{equation}
We show $D\left( E\right) $ in Fig. \ref{Fig:DOS} for various values of $\ell E_z$ and $M$. Notice that the DOS depends heavily on the spin and valley  index of the electrons. In Fig. \ref{Fig:DOS}(c) and (d) for example, the total DOS is constant near zero energy, but it nonetheless consists of different portions of spin-up and spin-down  states for each energy. 

\section{Polarization in silicene}\label{Sec:Polarization}
To assess electron-electron interactions in the RPA, we first need to determine the polarization:\cite{Kotov2008, Kotov2008a, Kotov2012, Pyatkovskiy2009}
\begin{equation}
\Pi ^{0}\left( \mathbf{q},\omega \right) =\sum_{\eta = \pm 1}  \int \frac{d\nu d^{2}k}{\left( 2\pi \right) ^{3}}\Tr\left[ G_{\eta}^{0}\left( \mathbf{k},\nu \right) G_{\eta}^{0}\left( \mathbf{k+q},\nu +\omega \right) \right],  \label{PolFun}
\end{equation}
where the summation over $\eta$ corresponds to  different valleys and $G_{\eta}^{0}\left( \mathbf{k},\nu \right) $ is the Green's function of the non-interacting particle near the $K_{\eta}$ valley. For a finite-mass Dirac Hamiltonian, such as the one shown in Eq. (\ref{Eq:Hamiltonian2By2}), the Green's function is given by\cite{Kotov2008, Kotov2008a} 
\begin{eqnarray}
\hspace*{-0.6cm}G_{\eta s_{z}}^{0}\left( \mathbf{k},\omega \right)  &=&\left( \hbar \omega +\mu _{0}-H_{\eta}^{s_{z}}\right) ^{-1}  \notag \\
&=&\hspace*{-0.2cm}\frac{\hbar \omega +\mu _{0}-2 s_{z}M+H_{\eta}^{s_{z}}}{\left( \hbar \omega + \mu _{0}-s_{z}M\right)^{2} - (\hbar v_{F}k)^{2}-\Delta _{\eta s_{z}}^{2}},
\end{eqnarray}
where we have omitted the identity matrices for the sake of brevity and $k$ is the magnitude of $\mathbf{k}$. Notice that the dependence of the Green's function on the spin quantum number $s_{z}$ is twofold: on the one hand it changes the Fermi level $\mu _{0}$ to $\mu _{s_{z}}=\mu_{0} -s_{z}M$ and on the other it influences the gap given by $\Delta _{\eta s_{z}}$ defined in Eq (\ref{Eq:Gap}).

We consider only the reduced Hamiltonian (\ref{Eq:Hamiltonian2By2}). For the complete $4\times 4$ Hamiltonian, we obtain both Green's functions along the diagonal, so 
\begin{equation}
G_{\eta}^{0}\left( \mathbf{k},\nu \right) =\left[ 
\begin{array}{cc}
G_{\eta, +1}^{0}\left( \mathbf{k},\nu \right) & 0 \\ 
0 & G_{\eta, -1}^{0}\left( \mathbf{k},\nu \right)
\end{array}
\right] .
\label{Eq:GreensFunction}
\end{equation}

We can readily calculate the trace in Eq. $( \ref{PolFun}) $ and write the total polarization as 
\begin{equation}
\Pi ^{0}\left( \mathbf{q},\omega \right) =\sum_{\eta = \pm 1}\sum_{s_{z}=\pm 1}\Pi_{\eta s_{z}}^{0}\left( \mathbf{q},\omega \right),
\label{Eq:TotalPolarization}
\end{equation}
with the spin- and valley-dependent polarization,  obtained in the manner of Ref.  \onlinecite{Pyatkovskiy2009}, given by
\begin{eqnarray}
\Pi _{\eta s_{z}}^{0}\left( \mathbf{q},\omega \right) &=&\frac{1}{2}\int \frac{d^{2}k}{\left( 2\pi \right) ^{2}}\sum_{\lambda \lambda ^{\prime }=\pm 1}f_{\eta s_{z}}^{\lambda \lambda ^{\prime }}\left( \mathbf{k,q}\right)  \notag \\
&\times& \frac{n_{F}^{s_{z}}\left( \lambda E_{\mathbf{k}}^{\eta s_{z}}\right)-n_{F}^{s_{z}}\left( \lambda ^{\prime }E_{\mathbf{k+q}}^{\eta s_{z}}\right) }{\hbar \omega +\lambda ^{\prime}E_{\mathbf{k}}^{\eta s_{z}}-\lambda E_{\mathbf{k+q}}^{\eta s_{z}}+i\delta }; \label{SpinDepPola}
\end{eqnarray}
here $\delta$ is an infinitesimal positive quantity, $n_{F}^{s_{z}}$ is the Fermi-Dirac distribution with a spin-dependent Fermi level $\mu _{s_{z}}$, $E_{\mathbf{q}}^{\eta s_{z}} =[\hbar^2 v_{F}^2q^{2} +\Delta_{\eta s_{z}}^{2}]^{1/2}$, and $f_{\eta s_{z}}^{\lambda \lambda ^{\prime }}$ the structure factor 
\begin{equation}
f_{\eta s_{z}}^{\lambda \lambda ^{\prime }}\left( \mathbf{k,q}\right) =\left[1+\lambda \lambda ^{\prime }\frac{\mathbf{k}\left( \mathbf{k+q}\right) +\Delta_{\eta s_{z}}^{2}}{E_{\mathbf{q}}^{\eta s_{z}}E_{\mathbf{k+q}}^{\eta s_{z}}}\right].
\label{Eq:StructureFactor}
\end{equation}
Equation (\ref{Eq:StructureFactor}) expresses the pseudo spinorial character of  electrons in silicene and will prevent two oppositely moving ones from interacting if the gap $\Delta_{\eta s_{z}}$ is zero. This factor is typical for graphene-like 2D systems such as silicene.

Since transitions between  different spin and valley states can be neglected, as previously motivated, the resulting polarization is a sum of independent systems. These systems are analogous to gapped graphene\cite{Pyatkovskiy2009}  but in each of them one can change the size of the gap separately by varying the  field $E_z$ and the Fermi level can be tuned by changing the  field $M$. The spin and valley components are influenced by these fields in  different ways. 

The contributions to the polarization from Eq. $\left(\ref{SpinDepPola}\right) $ can be written as a sum of three parts\cite{Wunsch2006, Hwang2007a} 
\begin{equation}
\Pi _{\eta s_{z}}^{0}=-\Pi_{\eta s_{z},\infty}^{-} +\Pi_{\eta s_{z},\mu _{s_{z}}}^{+} +\Pi_{\eta s_{z},\mu _{s_{z}}}^{-},
\end{equation}
where $\Pi_{\eta s_{z},\mu}^{\pm }=\Pi_{\eta s_{z},\mu}^{\pm }\left( \mathbf{q},\omega \right) $ stands for interband ($-$) or intraband ($+$) contributions  with Fermi level $\mu$. The last two terms contribute only for $\mu _{s_{z}}>\Delta_{\eta s_{z}}$. Note that because the Fermi level $\mu_{0}$ is fixed, while the spectrum is displaced or deformed due to the  field $E_z$ or $M$, it is possible that this inequality is reversed for one valley spin state while it still holds for the others. In such a case only the vacuum polarization $\Pi_{\eta s_{z},\infty }^{-}$ contributes to the polarization of the respective  state. Different possible situations are depicted in Fig. \ref{BandPlot}  in which the Fermi level is shown as a black dotted line. In Figs. \ref{BandPlot} (a) and (b) the Fermi level lies in the conduction band of all spin and valley type electrons such that intraband transitions are possible. In Fig. \ref{BandPlot} (c), however, this is not the case for  spin-up electrons near the $K_{-}$ valley. The combination of the $M$ and $E_{z}$ fields is such that the Fermi level lies in the band gap of this type of electrons and therefore intraband transitions are excluded in this case. In Fig. \ref{BandPlot} (d) the two valleys are equivalent as discussed earlier, but the Fermi level lies in the gap of the spin-up  electrons and excludes intraband transitions for them.

Analytical solutions have been found for both the vacuum polarization\cite{Kotov2008a, Pyatkovskiy2009, Son2007} and the polarization in graphene with a partly filled conduction band \cite{Pyatkovskiy2009}. The polarization for $\mu_{s_{z}}>\Delta _{\eta s_{z}}$ can be written as
\begin{equation}
\frac{\Pi _{\eta s_{z},\mu _{0}}^{0}\left( \mathbf{q},\omega \right) }{D_{0}\left( \mu _{0}\right) }=-\left[ \mu _{s_{z}}^{\prime }-\frac{
q^{\prime 2}}{8\,a_{q^\prime} } 
F_{\eta s_{z}}\left( \mathbf{q},\omega \right) \right] ,
\label{Eq:AnalyticalPolFull}
\end{equation}
where $a^2_{q^\prime}= \left\vert q^{\prime 2}-\omega^{\prime 2}\right\vert$, $q^{\prime } = \hbar v_{F}q/\mu _{0}$, $\omega ^{\prime } = \hbar \omega /\mu _{0}$, and $\mu _{s_{z}}^{\prime } = \mu _{s_{z}}/\mu _{0}$.  Further, $D_{0}\left( E\right) =\left\vert E\right\vert /[2 \pi \hbar ^{2}v_{F}^{2}]$ is the density of states for one spin component near one Dirac point if $E>\Delta_{\eta s_{z}}$ and $F_{\eta s_{z}}$  a dimensionless function that acquires different values in different parts of the $\left( q,\omega \right) \,$\ plane as shown in appendix \ref{Sec:Appendix}. 

On the other hand, the vacuum polarization  reads 
\begin{equation}
\frac{\Pi _{s_{z}, \text{vac}}^{0}\left( \mathbf{q},\omega \right) }{D_{0}\left( \mu_{0}\right) }=-\frac{q^{\prime 2}}{\,a_{q'}}\left[ \frac{\Delta _{\eta s_{z}}^{\prime }}{2 \, a_{q'}} +F_{\eta s_{z}, \text{vac}}\left( \mathbf{q},\omega \right) \right] ,
\label{Eq:AnalyticalPolVac}
\end{equation}
where $a^2_{q'}=|q'^2-\omega^2|$, $\Delta _{\eta s_{z}}^{\prime }=\Delta _{\eta s_{z}}/\mu _{0}$. $F_{\eta s_{z},\text{vac}}$ is a dimensionless function defined in appendix \ref{Sec:Appendix}. Combining these results, the total polarization is given by
\begin{equation}
\Pi _{\eta s_{z}}^{0}=\theta \left( \mu _{s_{z}}-\Delta _{\eta s_{z}}\right) \Pi_{\eta s_{z},\mu _{0}}^{0} + \theta \left( \Delta _{\eta s_{z}}-\mu _{s_{z}}\right) \Pi_{\eta s_{z},\text{vac}}^{0}.
\end{equation}

\begin{figure}[tb]
\centering
\includegraphics[width= 8.5cm]{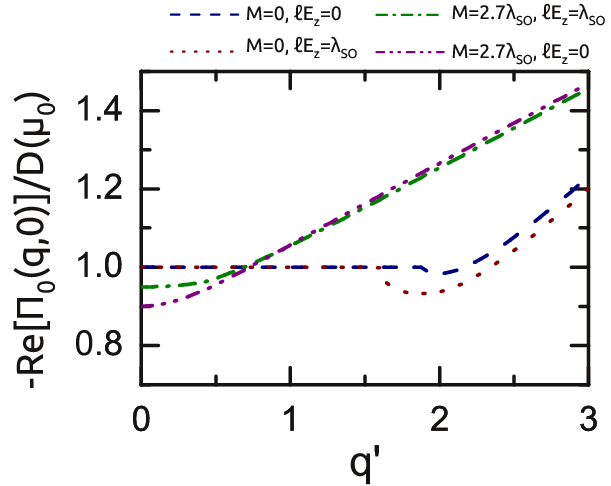}
\vspace*{-0.5cm}
\caption{(Colour online) Static polarization  versus wave vector $q^\prime$ for different values of the fields $E_{z}$ and $M$. The  Fermi level is $\mu_{0} =13$ meV and the SOC $\lambda_{SO} = 0.3\mu_{0} $. The blue dashed curve is for $E_{z}=M=0$,and the red dotted one for the critical field $E_{z}=E_{c}=\lambda_{SO}/\ell$ and $M=0$. Further, the green dash-dotted curve is for $M=2.7 \lambda_{SO}$ and  $\ell E_{z}=\lambda_{SO}$, and the purple dash-dot-dotted one  for $\ell E_{z}=0$ and $M=2.7 \lambda_{SO}$.}
\label{Fig:StaticPol}
\end{figure}

The complete expression of the dynamical polarization is given in appendix \ref{Sec:Appendix}. The static one can be written as
\begin{eqnarray}
\hspace*{-0.4cm}\frac{\Pi _{\eta s_{z},\mu _{0}}^{0}\left( \mathbf{q},0\right) }{D_{0}\left( \mu_{0}\right) } &=&-\Big[ \mu _{s_{z}}^{\prime }-\frac{1}{4}\theta \left( q-2q_{F,s_{z}}\right) \nonumber \\*
&\times & \big[ \frac{2\mu _{s_{z}}^{\prime }b_{q^\prime}}{ q^{\prime}} -\frac{c_{q^\prime}}{q^{\prime }}\arctan \frac{b_{q^\prime}}{ 2\mu _{s_{z}}^{\prime }} \big]\Big], \nonumber \\*
\hspace*{-0.3cm}\frac{\Pi _{\eta s_{z},\text{vac}}^{0}\left( \mathbf{q},0\right) }{D_{0}\left( \mu_{0}\right) } &=&-\frac{1}{2}\Big[ \Delta _{\eta s_{z}}^{\prime }+\frac{c_{q^\prime}^2}{2 q^{\prime }}\arcsin \frac{q^{\prime }}{c_{q^\prime}} 
\Big] ,
\label{Eq:PolStatic}
\end{eqnarray}
where $b_{q^\prime} =[q^{\prime 2}-4q_{F,s_{z}}^{\prime 2}]^{1/2}$ and $c_{q^\prime}=[q^{\prime 2}-4\Delta_{s_z}^{\prime 2}]^{1/2}$; we used the definition of the Fermi wave vector $q_{F,\eta s_{z}} = \hbar v_{F}[\mu^{2}_{s_{z}}-\Delta^{2}_{\eta s_{z}}]^{1/2}$.  For $M=0$ Eq. (\ref{Eq:PolStatic}) coincides with the  result of Ref. \onlinecite{Tabert2014a} and for $M=E_z=0$ and $\lambda_{SO}=\Delta$ with that of Ref. \onlinecite{Pyatkovskiy2009}. In Fig. \ref{Fig:StaticPol} we show the static polarization for various values of the  fields $E_z$ and $M$. This figure shows that as long as  all spin and valley type electrons contribute, for low $q$, the static polarization equals the DOS at the Fermi level, $D(\mu_{0}$), which is the case for the blue dashed and red dotted curves. However, when this is not the case, as for the other two curves, the polarization decreases because in this case, for some spin and valley type electrons only interband contributions are  allowed. 

Furthermore, one can approximate the polarization at small energies and momenta, $\hbar v_{F} k \ll \hbar \omega \ll \mu $ as\cite{Pyatkovskiy2009}

\begin{equation}
\hspace*{-0.4cm}\frac{\Pi _{\eta s_{z}}^{0}\left( \mathbf{q},\omega \right) }{D_{0}\left( \mu_{0}\right) } =  \frac{q^{\prime 2}\mu _{s_{z}}^{\prime }}{2\omega ^{\prime 2}}\left( 1-\frac{\Delta _{\eta s_{z}}^{\prime 2}}{\mu _{s_{z}}^{\prime 2}}\right) \theta \left( \Delta _{\eta s_{z}}^{\prime }-\mu _{s_{z}}^{\prime
}\right), 
\label{Eq:PlasmonApprox}
\end{equation}
which is valid for all values of $\mu_{s_z}$ and $\Delta_{\eta s_z}$. 

\subsection{Particle-hole excitation spectrum (PHES)}

The PHES is the region in the $\left(q,\omega \right) $-plane where it is possible for a photon with energy $\hbar \omega $ and momentum $\hbar q$ to excite an electron-hole pair. This ability for pair creation is embodied in the polarization, in  the regions where it has a non-zero imaginary part. The PHES can be divided into two disjunct regions where inter- and intraband electron-hole pair formation is possible. These regions are located above and below the $\omega^{\prime} = q^{\prime}$ line, respectively. Because the fields $E_z$ and $M$ change the structure of the bands and the position of the Fermi level, the PHES is also subject to changes in their values.  Since the valley and spin of the electrons determine how the fields affect their dispersion and Fermi level, the PHES can be different for electrons with different spin or valley index.

\begin{figure}[tb]
\centering
\includegraphics[width= 8.5cm]{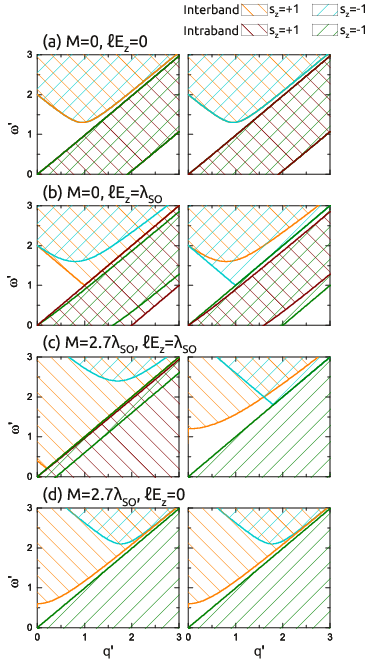}
\vspace*{-0.2cm}
\caption{(Colour online) PHES of silicene for different values of the  fields  $E_{z}$ and $M$ as specified.  The left column is for the $K_{+}$ valley, the right one for the $K_{-}$ valley. The different hatched regions correspond to the spin type of the electrons as indicated at the top of the figure. The different colours refer to the inter- or intraband PHES.  The SOC is 
$\lambda_{SO}=0.3\mu_0.$}

\label{PHESPlot}
\end{figure}

In Fig. \ref{PHESPlot} we show the PHES of silicene for different values of $E_z$ and $M$ for which the energy spectra are shown in Fig. \ref{BandPlot}. The left column corresponds to electrons in the $K_{+}$ valley and the right one to the $K_{-}$ valley. 

Figure \ref{PHESPlot}(a) shows that in the absence of the fields both valleys and spins are treated equally and the PHES is the same. When a finite electric field is applied, the gap is changed for each spin and the PHES of each spin near the same Dirac point is different. In the other valley, however, the effect of the electric field has the opposite effect such that the PHES is interchanged between both spin types. Figure \ref{PHESPlot}(b) shows the PHES when the critical electric field $E_{c}$ is applied.  Since for this field one of the two spin states has a gapless dispersion, we obtain the PHES of graphene for the spin-up state near the $K_{+}$ valley and the spin down-state near the $K_{-}$ valley. The situation is reversed if $E_{c}$ points in the opposite direction. For $M=0$ the effect of the electric field on the PHES was  investigated in detail in Ref. \onlinecite{Tabert2014a} 

Changing the exchange field $M$ also affects the PHES because the Fermi level of the two spin components is not the same. If the Fermi level of one component is moved inside the bandgap of the corresponding spectrum, the intraband region completely disappears because there are no electrons left in the conduction band.  This situation is depicted in Figs. \ref{PHESPlot}(c) and (d) where a finite field $M $ is considered. As seen in Fig. \ref{BandPlot}(c), for which the same fields are considered as Fig. \ref{PHESPlot}(c), the Fermi level $\mu_{+1}$ of the spin-up electrons is displaced inside the gap near the $K_{-}$ valley and therefore the intraband PHES is absent in this case. The PHES of  different spin and valley type electrons differs strongly, so spin and valley polarization is expected because there are regions in the $(q,\omega)$ plane where only one  spin and valley type of electrons can create  an electron-hole pair.

If the field $E_z$ is absent but the $M$ one is finite, the spin symmetry is broken but the valleys remain equivalent.  Figure \ref{PHESPlot}(d)  shows that the corresponding PHES is strongly spin dependent and  spin polarization is possible.

In the region between the intra- and interband PHES, it is not possible to excite a particle-hole pair. Therefore, in this region it is possible to  excite stable plasmons. Inside the PHES, the plasmons do have a finite lifetime because they can decay into electron-hole pairs. Notice that  the PHES  has a spin and valley texture, that is, in some parts of the $(q,\omega)$ plane  pair formation is allowed for only one spin and one valley component, denoted by $\{s_z, \eta\}$,  but that there are also regions shared with different  $\{s^\prime_z, \eta^\prime\}$ components. When  pair formation is allowed for only one spin and valley type, it is possible to excite plasmons that contain only one specific spin or valley type of electrons and refer to them as spin- and valley-polarized plasmons.

\subsection{Plasmons in the RPA}

\begin{figure*}[tb]
\centering
\includegraphics[width= 17 cm, height =7 cm]{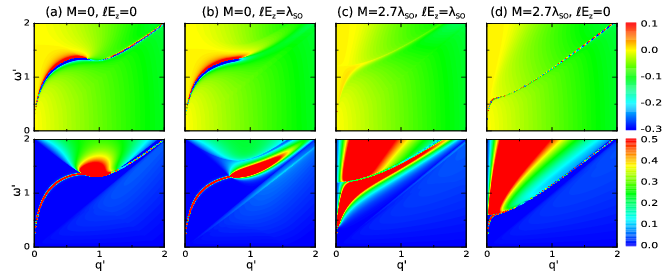}
\vspace*{-0.2cm}
\caption{(Colour online) Top row: real part of the  RPA polarization  $\Pi^{\prime}(\textbf{q},\omega)$ . Bottom row: absorption. 
For clarity we added a small imaginary part to $\omega$, such that $\omega\to\omega + I \beta$ with $\beta = 10^{-4}$. The columns correspond to various values of the  electric and exchange fields, as specified,  that are the same as those used in Figs. 1, 2, and 3. The SOC strength is $\lambda_{SO}=0.3\mu_{0}$.}
\label{Fig:2DPlot}
\end{figure*}

Using the expression found in Eq. $\left( \ref{Eq:TotalPolarization}\right) $, we can find higher-order contributions, within the RPA, using \cite{Goerbig2011, GiulianiBook, Ando1982}
\begin{equation}
\Pi \left( \mathbf{q},\omega \right) =\Pi ^{0}\left( \mathbf{q},\omega
\right)/ \varepsilon \left( \mathbf{q},\omega \right), 
\end{equation}
where the RPA dielectric function is given by 
\begin{equation}
\varepsilon\left(\mathbf{q},\omega\right) = 1 - v_{q}\Pi ^{0},
\end{equation}
with $v_{q}=e^{2}/2\kappa q$ the 2D Fourier transform of the Coulomb potential with an effective dielectric constant $\kappa$ that takes into account the  medium surrounding the silicene sample. In this study we consider free-standing silicene and we therefore take $\kappa=\epsilon_{0}$, the permittivity of vacuum. In Fig. \ref{Fig:2DPlot} we plot the real part of the RPA polarization function on the top row. 

Using the expressions in Eqs. (\ref{Eq:AnalyticalPolFull}) and (\ref{Eq:AnalyticalPolVac}), we can write the dielectric function as
\begin{equation}
\varepsilon\left(\mathbf{q},\omega\right) = 1 - (4\alpha/q^{\prime}) \Pi^{\prime, 0},
\end{equation}
where $\Pi^{\prime, 0} = \Pi^{ 0}/D(\mu_0)$  with $D(\mu_{0})=2\mu_{0}/\pi \hbar^{2} v_{F}^{2}$ and $\alpha=e^{2}/4 \pi \kappa \hbar v_{F}$ the fine-structure constant. The value of $\alpha$ depends on the material by virtue of the Fermi velocity $v_{F}$ and on the dielectric constant $\kappa$. For instance, in free-standing silicene we have $\alpha \approx 4$ while for silicene on a substrate, with $\kappa = 4$ (e.g. $SiO_{2}$), $\alpha$ decreases to $\alpha \approx 1$. In this work we consider only free-standing silicene. A new procedure to fabricate it, by  intercalating it between two graphene layers, was recently proposed,\cite{Berdiyorov2014,meh} and opens the way for its experimental realization. 

The dielectric function determines the amount of absorbed radiation through the relation 
\begin{equation}
A\left(\mathbf{q},\omega\right) = -\Im [1/\varepsilon\left(\mathbf{q},\omega\right)].
\end{equation}
In Fig. \ref{Fig:2DPlot} we show the absorption on the  bottom row.  The absorption is strong inside the interband PHES because here particle-hole pair formation is possible. There is, however, also a pronounced absorption curve outside the PHES. Here the radiation is not absorbed by pair formation but by plasmons. 

\begin{figure*}[t!]
\centering
\includegraphics[width= 17 cm,  height =9.5 cm]{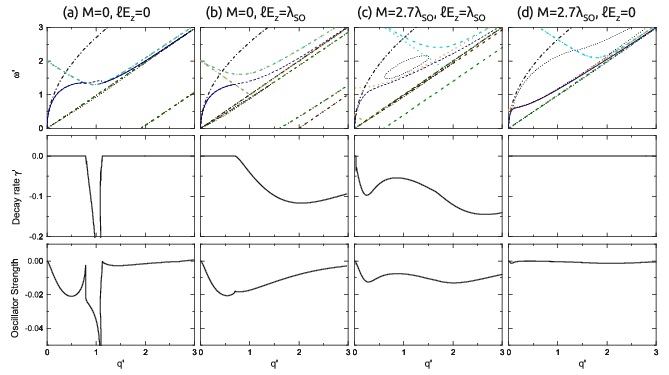}
\vspace*{-0.4cm}
\caption{(Colour online) Top row:  plasmon dispersion in silicene. The blue solid curve corresponds to plasmons with zero decay rate, the blue dashed curves are the roots of the real part of the dielectric function  that are the prolonged versions of the stable branches, the blue dotted curves are the other roots of the real part of $\varepsilon$, the dash-dotted curve is the low-momentum approximation.  The other coloured curves show the borders of the PHES of  different spin and valley types as shown in Fig. \ref{PHESPlot} with the same colour code. The dashed curves are for the $K_{+}$ valley and the dotted ones for the $K_{-}$ valley. Second and third  rows: decay rate and oscillator strength, respectively, of the plasmon branches shown in blue on the first row. Each column corresponds to various values of the electric and exchange fields as specified.}
\label{Fig:PlasPlot}
\end{figure*}

Plasmons in the RPA can be found from the zeros of the dielectric function $\varepsilon\left( \mathbf{q},\omega_{p} + i \gamma \right) $, where $\omega_{p}$ is the plasma frequency and $\gamma$ the decay rate of the plasmon. Following the standard procedure \cite{Wunsch2006, Hwang2007a, Pyatkovskiy2009} we first look for the zeros of the real part of the dielectric function  $\Re \varepsilon \left( \mathbf{q},\omega \right) $.  Later on, we will include the finite lifetime. Note that these plasmons are exact and not damped outside the PHES since the imaginary part of the polarization is zero. 

The plasmon dispersion is found by solving the equation $\Re \varepsilon \left( \mathbf{q},\omega \right) =0$ numerically. However, for small energies and momenta we can  find a near analytical solution upon using Eq. (\ref{Eq:PlasmonApprox}). The approximate result is
\begin{equation}
\hspace*{-0.1cm}\omega _{p}^{\prime 2} \approx \frac{\alpha q^{\prime }}{8}\sum_{\eta, s_{z}=\pm
1}\mu _{s_{z}}^{\prime }\Big( 1-\Delta _{\eta s_{z}}^{\prime 2}/\mu
_{s_{z}}^{\prime 2}\Big) \theta ( \Delta _{\eta s_{z}}^{\prime }-\mu
_{s_{z}}^{\prime }).
\label{Eq:ApprPlasmon}
\end{equation}
Note that we obtain again a $\sqrt{q}$ behaviour. This is typical for plasmons in 2D systems.\cite{Krstajic2013a}

On the first row of Fig. \ref{Fig:PlasPlot} we show the plasmon branches for several configurations of the electric and exchange fields. The solid curves correspond to the plasmons with zero decay rate while the dashed curves are the zeroes of the real part of $\varepsilon$. The plasmon dispersion given by Eq. (\ref{Eq:ApprPlasmon}) is shown by dash-dotted curves.

A more detailed study of the effect of the electric field on plasmons in silicene can be found in Ref. \onlinecite{Tabert2014a}; our results  reduce to those of this study for $M=0$.

\subsection{Oscillator strength and decay rate}

In this part we calculate the oscillator strength and the decay rate which are related to the robustness of the plasmon. The decay rate determines how fast the plasmon decays after it has been excited. It is given by\cite{Wunsch2006, FetterBook} 
\begin{equation}
\gamma =\frac{\Im \Pi ^{0}\left( \mathbf{q},\omega _{p}\left( \mathbf{q}\right) \right) }{\left[ \left( \partial /\partial \omega \right) \Re \Pi^{0}\left( \mathbf{q},\omega \right) \right] _{\omega =\omega _{p}\left( \mathbf{q}\right) }},
\end{equation}
where $\omega _{p}\left( \mathbf{q}\right)$ is the plasmon frequency for wave vector \textbf{q}. In Fig. \ref{Fig:PlasPlot} on the second row we show the decay rate as a function of the plasmon momentum. 

The amount of absorbed radiation by the plasmon is determined by the oscillator strength. In the RPA the imaginary part of the dynamical polarization near the plasmon branch is given by\cite{GiulianiBook}

\begin{equation}
\Im \Pi \left( \mathbf{q},\omega \rightarrow \omega_{p}\right) = O\left(\mathbf{q}\right) \delta \left(\omega -\omega _{p} \right) .
\end{equation}

and the oscillator strength $O\left(\mathbf{q}\right)$ by 

\begin{equation}
O\left(\mathbf{q}\right) = -\frac{\pi\,\Re \Pi ^{0}\left( \mathbf{q},\omega _{p} \right) }{v_{q}\left\vert \partial \Re \Pi ^{0}\left( \mathbf{q},\omega \right) /\partial \omega ] \right\vert_{\omega =\omega _{p}}}
\end{equation}

In Fig. \ref{Fig:PlasPlot} we show the oscillator strength for various systems as a function of the wave vector on the third row. 

\section{Numerical results}\label{Sec:NumericalRes}

The results shown in Figs. \ref{Fig:2DPlot} and \ref{Fig:PlasPlot} correspond to four systems  to which different electric and exchange fields are applied. These four different combinations of $E_{z}$ and $M$ correspond to the values used earlier in Figs. (\ref{BandPlot})-(\ref{PHESPlot}). This allows us to demonstrate the different types of plasmons that can be excited in silicene. In all cases we use $\lambda_{SO}=3.9$ meV.

In the first column we show results for  a 
Fermi level $\mu_{0}\approx 3 \lambda_{SO}$ but  no external fields present. Here both spin types behave the same and we obtain a plasmon branch that is similar to that of gapped graphene\cite{Pyatkovskiy2009}. Note, however, that because the fine structure constant $\alpha$ is twice that of  free-standing graphene, it is interrupted by the interband PHES which is accompanied by a strong absorption. Inside the PHES, the decay rate increases very quickly showing the instability of the plasmon at that point. The oscillator strength of the small $q$ plasmon branch is larger than that of the second part of the stable branch. The first part is therefore more pronounced than the second one.

In the second column  $E_{z}$ is the critical field, $\ell E_{z} = \lambda_{SO}$, and the exchange field is zero. This particular example is especially interesting because the spectrum of the $s_{z}=+1$ electrons is linear and gapless while that of the $s_{z}=-1$ electrons has a large gap  near the $K_{+}$ valley. The corresponding plasmons have a dispersion that is similar to that of graphene, but which is interrupted by the PHES of the $K_{+}$ spin-up electrons. The absorption shows that at this point the plasmon can decay into spin-up electron hole pairs near the $K_{+}$ valley. This is supported by the increase in the decay rate,  but note that this  rate is much smaller than that for the first case because only one spin type per valley can induce pair formation. The oscillator strength of the plasmon outside the PHES is similar to that of the field-free case, but diminishes inside the PHES. Although  the PHES  and the plasmons in the two valleys  depend on the spin type, they are compensated in the other valley. Therefore, in this case the plasmons are not spin polarized. 

In the third column the electric and exchange fields are finite.  Because of the combined effect of both fields, the PHES acquires a spin and valley texture. For very small $q$ the plasmon branch is stable but it quickly encounters the border of the $K_{+}$ spin-up PHES where the decay rate increases. The branch continues under the border of the $K_{-}$ spin-up PHES but the strong absorption signals its presence. The crossing of the border of the $K_{-}$ spin-down PHES results in a shortening of the lifetime because the plasmons can decay into two different types of electrons, the spin-up type in the $K_{+}$ valley and the spin down one in the $K_{-}$ valley. Thus, in this case we obtain spin- and valley-polarized plasmons. Notice an additional feature, the oval dotted curve; this is the remnant of the plasmon branch of spin-down electrons.

The fourth column applies to the case where the electric field is absent but a very large exchange field is applied, $M \approx 2.7 \lambda_{SO}$. In this situation the dispersions of both spin types are shifted with respect to each other in such a way that the Fermi level lies in the gap of the spin-up electrons while it is situated in the conduction band of the spin-down electrons. For spin-up electrons the intraband PHES therefore vanishes and the interband PHES is shifted to lower energies.  Here there is an undamped plasmon branch that follows the border of the interband PHES of the spin-up electrons.  Despite its stability, the oscillator strength and absorption show that it is not very pronounced. There exists, however, also a strongly damped branch that lies inside the spin-up interband PHES, shown by the dotted curve, which is the prolonged version of the undamped plasmon and that showing the same $\sqrt{q}$ dependence of the stable plasmon for low $q$. This branch  indicates a spin-down plasmon but one which can decay very quickly into spin-up electron-hole pairs leading to a strong absorption in this region  as shown in Fig. \ref{Fig:2DPlot}(d). This plasmon lies outside the PHES and is therefore neither spin- nor valley-polarized. However, the plasmon-like branch inside the spin-up PHES does resemble a spin-polarized system but with a very short lifetime. 

\section{Concluding remarks and outlook}\label{Sec:Concl}

We investigated how electric ($E_z$) and exchange ($M$) fields can be used to tune the plasmonic response of the electron gas in silicene. These fields affect the PHES of electrons with opposite spin and valley indices in  different ways  giving the PHES a spin and valley texture and thus leading to spin- and valley-polarized plasmons. Their lifetime is, however, finite because they can easily decay into  electron-hole pairs with different spin or valley indices. Further, the  field $M$ affects strongly the oscillator strength. The undamped plasmon that remains has a negligible strength and  is therefore not expected to show up in experiments.  If the Fermi level lies in the gap of a spin in one valley, the intraband region of the corresponding PHES disappears. For zero $E_z$ and  finite $M$ the spin symmetry is broken  and spin polarization is possible. 

We found a low-momentum plasmon dispersion that has the typical 2D $\sqrt{q}$ behaviour. At higher momentum, the plasmon dispersion differs from this $\sqrt{q}$ dependence.  However, a strong  field $M$ can induce plasmons with finite lifetime of one spin type for which the dispersion follows closely the approximate plasmon branch.

The results reported in this work pertain to free-standing silicene since we used the permittivity $\epsilon_{0}$ of the vacuum. For silicene on a substrate the permittivity will be different and the results will be modified quantitatively. The plasmon branches will be curved downward and the PHES can be avoided yielding plasmons with larger oscillator strength. 

We evaluated the plasmon dispersion within the framework of the RPA  which is valid for high electron densities. For lower densities further work is necessary to obtain more accurate results for silicene's plasmonic properties.

The predicted spin-and valley-polarization of plasmons is a consequence of a mechanism similar to  that responsible for  directional filtering of plasmons in a 2D electron gas.\cite{Badalyan2009} In this case the anisotropy of the electron dispersion renders the PHES  anisotropic and  damps the plasmons in one direction more than in the other.

Recently a lot of experimental progress has been made in the detection of plasmons in 2D materials.\cite{Luo2013c} To overcome the mismatch of energy and momentum with light in free space, one uses a periodic diffractive grid, cuts the sheet into ribbons or other shapes, and uses an AFM tip to excite the electron liquid or uses a resonant metal antenna\cite{Antenna}. The appropriate experimental probe though will strongly depend on the setup used to stabilize the silicene crystal. We believe that at the current pace of experimental progress, in the creation of silicene and the probing of plasmons in 2D materials, the results presented in this paper will soon be tested experimentally.

\vspace*{-0.3cm}
\section*{Acknowledgments}
This work was supported by the European Science Foundation (ESF) under the EUROCORES Program Euro-GRAPHENE within the project CONGRAN, the Flemish Science Foundation (FWO-Vl) by an aspirant grant to BVD, the Methusalem Foundation of the Flemish Government, and by the Canadian NSERC Grant No. OGP0121756.

\appendix
\section{Analytical formulae}\label{Sec:Appendix}

The results for the polarization shown in Eqs. $\left(\ref{Eq:AnalyticalPolFull}\right)$ and $\left(\ref{Eq:AnalyticalPolVac}\right)$ each depend on a dimensionless function that determines the variation of the polarization in the $(q,\omega)$ plane. For the vacuum polarization this function is given in terms of the dimensionless variables $q^{\prime}$, $\omega^{\prime}$ and $\mu_{s_{z}}^{\prime}$ defined earlier. In the following, we will suppress the prime notation for simplicity and set $a_q^{2}=\omega^2-q^2$. Then\cite{Pyatkovskiy2009} 
\begin{eqnarray}
\nonumber
&&\hspace*{-0.5cm}F_{\eta s_{z},\text{vac}} =\frac{4\Delta_{\eta s_z}^{2}+a_q^{2}}{8\,a_q^{2}}\Big[ \theta( q + \omega) \arccos \frac{4\Delta_{\eta s_z}^{2}+a_q^{2}}{4\Delta_{\eta s_z}^{2}-a_q^{2}}+\theta ( \omega-q)\\*
&&\hspace*{-0.5cm}\,\,\times 
\ln \frac{\left(2\Delta_{s_{z}}+a_q \right) ^{2}}{\left \vert 4 \Delta_{\eta s_z}^{2}-a_q^{2} \right \vert }\Big]
+i\frac{\pi }{8}\Big( \frac{a_q^{2}-4\Delta_{\eta s_z}^{2}}{a_q^{2}}\Big)
\theta \big(a_q^{2}-4\Delta_{\eta s_z}^{2}\big).
\end{eqnarray}

With $y_\pm=(2\mu _{s_{z}} \pm\omega)/q$, $x_{\eta s_z} =[1+4\Delta _{\eta s_{z}}^{2}/(q^{2}-\omega^{2})]^{1/2}$, and $\bar{x}_{\eta s_z}^2 
= 2-x_{\eta s_{z}}^{2}$, the total polarization for $\mu_{s_{z}}>\Delta_{\eta s_{z}}$ is\cite{Pyatkovskiy2009}
\begin{gather*}
F_{\eta s_{z}}\left( q,\omega \right) = 
\begin{cases}
iG_{>}( y_-)-iG_{>}( y_+) &: 1A \\ 
G_{<}( y_-)\,\,  -iG_{>}( y_+) &: 2A \\ 
G_{<}( y_+)\,\,+G_{<}( y_-)&: 3A \\ 
G_{<}( y_-)\,\, -G_{<}( y_+)  &: 4A \\ 
G_{>}( y_+) \,\,-G_{>}( y_-)  &: 1B \\ 
G_{>}( y_+) \, \,+iG_{<}( y_-)  &: 2B \\ 
G_{>}( y_+) \,\,-G_{>}( -y_-)  -i\pi\, \bar{x}_{\eta s_z}^2 &: 3B \\ 
G_{>}(-y_-) \,\,+G_{>}( y_+)  -i\pi\, \bar{x}_{\eta s_z}^2 &: 4B \\ 
G_{0}( y_+) \,\,\,\, -G_{0}( y_-)  &: 5B
\end{cases}
\end{gather*}

In these expressions
\begin{eqnarray}
G_{<}(x) &=& x[x_{\eta s_z}^2-x^2]^{1/2} -\bar{x}_{\eta s_z}^2 \arccos \left( x/x_{\eta s_{z}}\right),  \\
G_{>}(x) &=&x [x^2-x_{\eta s_z}^2]^{1/2}- \bar{x}_{\eta s_z}^2 \arcosh\left( x/x_{\eta s_{z}}\right),  \\
G_{0}(x) &=&x[x^2-x_{\eta s_z}^2]^{1/2} -\bar{x}_{\eta s_z}^2 \arsinh\left( x/ix_{\eta s_{z}}\right) .
\end{eqnarray}

\begin{figure}[tb]
\centering
\includegraphics[width= 8.5cm]{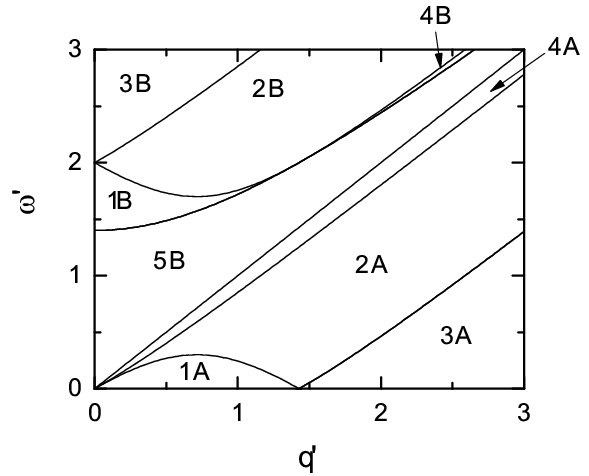}
\vspace*{-0.2cm}\caption{
Different regions of the ($q$-$\omega$) plane that are specified below. For this illustration we used $\mu_{s_{z}}=1$ and $\Delta_{\eta s_{z}} = 0.7 \mu_{s_{z}}$.}
\label{Fig:Regions}
\end{figure}

In this solution, we have subdivided the $(q,\omega)$ plane in several regions for which the solution is different. We show them in Fig. \ref{Fig:Regions}. Setting $q_\pm=q\pm q_{F,\eta s_z}$ the borders determining these regions are determined as

\begin{eqnarray*}
\hspace*{-0.65cm}
1A &:&\omega <\mu_{s_{z}}-[q_-^2 +\Delta _{\eta s_{z}}^{2}]^{1/2}\\
2A &:&\pm \mu _{s_{z}}\mp [q_-^2 +\Delta _{\eta s_{z}}^{2}]^{1/2} <\omega  <-\mu _{s_{z}}+[q_+^2 +\Delta _{\eta s_{z}}^{2}]^{1/2} \\
3A &:&\omega <-\mu _{s_{z}}+[q_-^2 +\Delta _{\eta s_{z}}^{2}]^{1/2}\\ 
4A &:&-\mu _{s_{z}}+[q_+^2 +\Delta _{\eta s_{z}}^{2}]^{1/2} <\omega < q \\
1B &:&q<2q_{F,\eta s_z}, [q^2 +4 \Delta _{\eta s_{z}}^{2}]^{1/2}< \omega <\mu _{s_{z}}+[q_-^2 +\Delta _{\eta s_{z}}^{2}]^{1/2}\\ 
2B &:&\mu _{s_{z}}+[q_-^2 +\Delta _{\eta s_{z}}^{2}]^{1/2} <\omega  <\mu _{s_{z}}+ [q_+^2 +\Delta _{\eta s_{z}}^{2}]^{1/2}\\ 
3B &:&\omega >\mu _{s_{z}}+[q_+^2 +\Delta _{\eta s_{z}}^{2}]^{1/2}\\ 
4B &:&q>2q_{F,\eta s_z}, [q^2 +4 \Delta _{\eta s_{z}}^{2}]^{1/2} <\omega  <\mu _{s_{z}}+[q_-^2 +\Delta _{\eta s_{z}}^{2}]^{1/2}\\ 
5B &:&q<\omega < [q^2 +4 \Delta _{\eta s_{z}}^{2}]^{1/2}. 
\end{eqnarray*}

Note that the PHES is defined as the region where the imaginary part of the polarization is non zero. This is the case in  regions $1A$, $2A$ for the intraband PHES and in  regions $2B$, $3B$, and $4B$ for the interband PHES of each valley spin state.

\end{document}